\RequirePackage{ifpdf}
\ifpdf % We are running pdfTeX in pdf mode
\documentclass[pdftex]{sigma}
\else
\documentclass{sigma}
\fi

\numberwithin{equation}{section}

\begin{document}

\renewcommand{\PaperNumber}{087}

\FirstPageHeading

\renewcommand{\thefootnote}{$\star$}

\ShortArticleName{Quantum Information from Graviton-Matter Gas}

\ArticleName{Quantum Information from Graviton-Matter
Gas\footnote{This paper is a contribution to the Proceedings of
the Seventh International Conference ``Symmetry in Nonlinear
Mathematical Physics'' (June 24--30, 2007, Kyiv, Ukraine). The
full collection is available at
\href{http://www.emis.de/journals/SIGMA/symmetry2007.html}{http://www.emis.de/journals/SIGMA/symmetry2007.html}}}

\Author{Lukasz-Andrzej GLINKA}

\AuthorNameForHeading{L.A. Glinka}

\Address{Bogoliubov Laboratory of Theoretical Physics, Joint
Institute for Nuclear Research,\\ 6 Joliot-Curie Str., 141980
Dubna, Moscow Region, Russia}
\Email{\href{mailto:glinka@theor.jinr.ru}{glinka@theor.jinr.ru},
\href{mailto:laglinka@gmail.com}{laglinka@gmail.com}}

\ArticleDates{Received May 16, 2007, in f\/inal form August 27,
2007; Published online September 04, 2007}

\Abstract{We present basics of conceptually new-type way for
explaining of the origin, evolution and current physical
properties of our Universe from the graviton-matter gas viewpoint.
Quantization method for the Friedmann--Lema\^{i}tre Universe based
on the canonical Hamilton equations of motion is proposed and
quantum information theory way to physics of the Universe is
showed. The current contribution from the graviton-matter gas
temperature in quintessence approximation is discussed.}

\Keywords{quantum cosmology; Friedmann Universe; nonequilibrium
thermodynamics; quantum information in cosmology}

\Classification{83F05; 62B10; 83C45; 83C47; 82C10; 70S05}

\section{Introduction}

Origin, evolution and physical picture of our Universe are one of
main areas of modern experimental as well as theoretical physics.
From the theoretical physics viewpoint it seems that all known
approaches, from the most popular inf\/lationary cosmo\-lo\-gy~\cite{M0} and loop quantum cosmo\-lo\-gy~\cite{B0} to new alternative
and compatible with experimental data conformal cosmo\-lo\-gy
scenarios~\cite{GP1, GP2, ZZP}, produce pictures of the Universe
incompatible with each other.

The most probable model for the Universe is the conformal f\/lat
Friedmann--Lema\^{i}tre spacetime. In this paper we present basics
of conceptually new way for explaining the origin, evolution and
current physical properties of our Universe. We found and develop
the quantum information theory approach to the structure and
physics of the Universe. As a model of the Universe we study the
classical Friedmann--Lema\^{i}tre spacetime. As the f\/irst
approximation we study the quintessence model.

By the well known Dirac method \cite{dir} we construct the
Hamiltonian approach to the Universe and as the solution of
classical constraints we obtain the Hubble law, that conf\/irms
experimental data. According to the Dirac approach we apply the
f\/irst quantization of classical primary constraints and as a
result we obtain the Wheeler--DeWitt equation for the Universe.
For correct quantum description of quantum gravity in this
cosmological model we propose the second quantization method for
the Wheeler--DeWitt equation by non-fockian distributions and by
that we obtain quantum description of gravity in terms of the
graviton-matter gas. By using the Bogoliubov transformation and
diagonalization of equations of motion we build the correct Fock
space formulation for quantum theory of the Universe. We show that
a~crucial role for the quantum theory of gravitation is played by
quantum ef\/fective phenomena in the graviton-matter gas, which in
case of the Friedmann--Lema\^{i}tre spacetime are superf\/luidity
phenomena. We f\/ind the f\/ield operator of the Universe. By
using the quantum information theory methods we calculate the
density matrix and entropy of the graviton-quintessence gas. We
predict a~relation between temperature and number of particles,
and evaluate equation of state for the gas.

In this paper we will use cosmological units system
$\hbar=c=k_B=\frac{8\pi G}{3}=1$, where $\hbar$ is the Dirac
constant, $c$ is the velocity of light, $k_B$ is the Boltzmann
constant, $e$ is the elementary charge and $G$ is the Newton
gravitational constant.

\section[Classical Friedmann-Lema\^{i}tre spacetime]{Classical Friedmann--Lema\^{i}tre spacetime}

\subsection[Conformal flat metric]{Conformal f\/lat metric}

As a cosmological model of the Universe we will study an exact
solution of the Einstein f\/ield equations of general relativity
\cite{einstein}, homogenous, f\/lat and isotropic expanding or
contracting spacetime founded and studied by A.A.~Friedmann
\cite{aaf} in the Einstein f\/ield equations context and by G.-H.
Lema\^{i}tre \cite{lem} in the Big Bang theory as the origin of
the Universe context. This model is characterized by the interval
\begin{gather}\label{3.1}
ds^2=dtdt-a^{2}(t)dx^idx^i,
\end{gather}
where $a(t)$ is known as the Friedmann conformal scale factor. The
spacetime interval (\ref{3.1}) can be transformed to conformal
f\/lat form by dif\/feomorphism of $t$. Friedmann \cite{aaf}
introduced change of time $t$ on the conformal time $\eta$ by
formula
\begin{gather}
\eta=\int_{t_0}^{t}\dfrac{dt'}{a(t')}.\label{3.2}
\end{gather}
This is a replacement of two dif\/feoinvariant times
$t\rightarrow\eta$. With using of the conformal time (\ref{3.2})
the interval (\ref{3.1}) reduces into the pseudo-Euclidean form
\[
ds^{2}=a^{2}(\eta)\left(d\eta d\eta-dx^idx^i\right).
\]

Conceptually new moment in the general relativity was introducing
by P.A.M.~Dirac \cite{dir} of the lapse function $N_d(x^0)$
def\/ined by the formula
\begin{gather}\label{eta}
d\eta=N_d(x^0)dx^0,
\end{gather}
where $x^0$ is time-coordinate as object of dif\/feomorphisms
\[
x^0\rightarrow \widetilde{x}^0=\widetilde{x}(x^0),
\]
introduced by Albert Einstein \cite{einstein} in the general
relativity context and developed by A.L.~Zelmanov \cite{zelmanov}
in cosmology context and by B.M.~Barbashov \emph{et al.}
\cite{barbashov} in dif\/feoinvariant Hamiltonian cosmological
perturbation theory.

\subsection{The Dirac Hamiltonian approach}

Recall that the Einstein--Hilbert general relativity with presence
of the matter f\/ields is described by the action found by David
Hilbert \cite{H}
\begin{gather}\label{2.1}
\mathcal{A}=\int{d^{4}x}\sqrt{-g}\left\{-\dfrac{1}{6}\mathcal{R}+\mathcal{L}_{M}\right\},
\end{gather}
where $g=\det{g_{\mu\nu}}$, $g_{\mu\nu}$ is the metric tensor of
the spacetime, $\mathcal{L}_{M}$ is the matter f\/ield Lagrangian
and $\mathcal{R}$ is the Ricci scalar (see for example
\cite{Mis,wein}). In cosmological considerations the
Lagrangian~$\mathcal{L}_M$ describes matter in the Universe. As
the Universe is classical object, the matter is characterized by
mean-f\/ield properties.

The Hilbert action (\ref{2.1}) calculated for the
Friedmann--Lema\^{i}tre metric (\ref{3.1}) is
\begin{gather}\label{3.5}
\mathcal{A}[a]=-V\int{dx^0}\left\{\dfrac{1}{N_d}\left(\dfrac{da}{dx^0}\right)^{2}+N_da^{4}\langle\mathcal{H}(x^0)\rangle\right\},
\end{gather}
where
\[
\langle\mathcal{H}(x^0)\rangle=\frac{1}{V}\int{d^{3}x}~\mathcal{H}_{M}(x^i,x^0),\qquad
V=\int{d^{3}}x<\infty,
\]
are the zeroth Fourier harmonic of the matter Hamiltonian and
spatial volume, respectively. We apply for the action (\ref{3.5})
the Hamiltonian reduction procedure. Firstly, we calculate the
canonical conjugate momentum corresponding to this action
\begin{gather}\label{3.6}
p_{a}=-\dfrac{2V}{N_d}\dfrac{da}{dx^0},
\end{gather}
and with use of this momentum the action (\ref{3.5}) becomes
\[
\mathcal{A}[a]=-V\int{dx^0}\left\{\dfrac{p_a^2}{4V^2}+a^{4}\langle\mathcal{H}(x^0)\rangle\right\}.
\]
From the Hamiltonian reduction viewpoint the reduced action has a
form
\begin{gather}\label{3.7}
\mathcal{A}[a]=\int{dx^0}\left\{p_a\dfrac{da}{dx^0}-\mathrm{H}(p_a,a)\right\},
\end{gather}
where the Hamiltonian $\mathrm{H}(p_a,a)$. By this we obtain
\[
\mathrm{H}(p_a,a)=N_d\left[-\dfrac{p_{a}^{2}}{4V}+V\langle\mathcal{H}(x^0)\rangle
a^{4}\right].
\]
According to the Dirac approach the action principle with respect
to the lapse function $N_{d}$ applied to the action (\ref{3.7})
produces Hamiltonian constraint equation. In the considered case
we have the constraint equation
\begin{gather}\label{3.8}
\dfrac{\delta\mathcal{A}[a]}{\delta
N_d}=0=-\dfrac{p_{a}^{2}}{4V}+V\langle\mathcal{H}(x^0)\rangle
a^{4}.
\end{gather}
We can resolve this classical constraint equation immediately. As
a result, we obtain
\[
\dfrac{a(t)}{a(t_0)}=\exp{\left\{\mathrm{sgn}(t-t_0)\int_{t_0}^{t}N_d(x^0)dx^0\sqrt{\langle\mathcal{H}(x^0)\rangle}\right\}},
\]
and it is the Hubble law.

From the other side the constraint equation (\ref{3.8}) expressed
in the Dirac conformal time has a solution
\begin{gather}\label{3.9}
p_{a}=-2V\dfrac{da}{d\eta}=\pm{\omega}_{a},
\end{gather}
and def\/ines values of the canonical conjugate momentum
(\ref{3.6}). In (\ref{3.9}) the quantity ${\omega}_{a}$ is time
dif\/feoinvariant variable
\begin{gather}\label{3.10}
{\omega}_{a}=2V\sqrt{\langle\mathcal{H}(\eta)\rangle}a^{2}(\eta).
\end{gather}
Equation (\ref{3.9}) produces the ordinary dif\/ferential equation
on $a(\eta)$
\[
-\dfrac{da}{d\eta}=\pm
\sqrt{\langle\mathcal{H}(\eta)\rangle}a^2(\eta).
\]
In this equation variables can be separated immediately and
elementary integration leads to the result
\begin{gather}\label{3.14}
a(\eta)=\dfrac{a(\eta_{0})}{1+z(\eta_{0};\eta)},
\end{gather}
where we have def\/ined the quantity
\begin{gather}\label{3.13}
z(\eta_{0};\eta)=a(\eta_0)\mathrm{sgn}(\eta-\eta_0)\int_{\eta_{0}}^{\eta}d\eta'\sqrt{\langle\mathcal{H}(\eta')\rangle}.
\end{gather}
The nature of $z(\eta_{0},\eta)$ can be understood if we rewrite
(\ref{3.13}) in the power series form \cite{wein, kolb}
\[
z(\eta_{0};\eta)=H_{0}(\eta-\eta_{0})+\left(1+\dfrac{q_{0}}{2}\right)H_{0}^{2}(\eta-\eta_{0})^{2}+\cdots.
\]
The quantity $z(\eta_{0},\eta)$ is nothing else than \emph{the
redshift}. The constants $H_{0}$ and $q_{0}$ are called the Hubble
parameter and the deceleration parameter
\begin{gather}
H_{0}=\sqrt{\langle\mathcal{H}(\eta_{0})\rangle}a(\eta_0),\label{3.14a1}\\
q_{0}=\dfrac{2}{H_0}\dfrac{\langle{\dot{\mathcal{H}}}(\eta_{0})\rangle}{\langle\mathcal{H}(\eta_{0})\rangle}-2.\label{3.14a2x}
\end{gather}
The result (\ref{3.14}) lies in agreement with experimental data,
this formula describes the Hubble law. It is clear that the Hubble
parameter (\ref{3.14a1}) and the deceleration parameter
(\ref{3.14a2x}) are dif\/feoinvariants.
\subsection{Quintessence}
We understand the quintessence as a kind of matter characterized
by constant energy -- the cosmological constant $\Lambda$. The
cosmological constant is equal to zeroth mode of the matter
Hamiltonian. By this way properties of the constant $\Lambda$ are
\[
\langle\mathcal{H}(\eta)\rangle=\langle\mathcal{H}(\eta_{0})\rangle=\Lambda,
\qquad
\langle\dot{\mathcal{H}}(\eta)\rangle=\langle\dot{\mathcal{H}}(\eta_{0})\rangle=0.
\]
In this approximation the Hubble constant (\ref{3.14a1}) and the
deceleration parameter (\ref{3.14a2x}) have a~simple form
\[
H_{0}=\Lambda^{1/2}a(\eta_0), \qquad q_{0}=-2,
\]
and the redshift is
\begin{gather}\label{a1}
z(\eta_{0};\eta)=H_0|\eta-\eta_{0}|.
\end{gather}
The solution of classical constraints (\ref{3.10}) for the
quintessence has a form
\[
p_a=\pm{\omega}_{a}(\eta)=\pm2V\Lambda^{1/2}a^{2}=\pm
{\omega}_{a}(\eta_0)\left(\dfrac{a(\eta)}{a(\eta_0)}\right)^2,
\]
where
\[
{\omega}_{a}(\eta_0)=2V\Lambda^{1/2}a^{2}(\eta_0)=2V\dfrac{H_0^2}{\sqrt{\Lambda}},
\]
is dif\/feoinvariant constant.

\section{Quantum gravity and collective phenomena}
In this section we quantize the Friedmann--Lema\^{i}tre spacetime
with quintessence. In contrast to hitherto existing approaches we
propose the procedure based on the Hamilton equations of motion
\begin{enumerate}\itemsep=0pt
\item By f\/irst quantization of classical constraints we obtain
the Wheeler--DeWitt equation for the wave function $\Psi$ of the
Universe, \item We treat the wave function $\Psi$ as a classical
f\/ield, and we construct classical f\/ield theory the canonical
Hamilton equations, \item We quantize the canonical Hamilton
equations by non-fockian distributions in the Fock space of
annihilation and creation operators, \item We apply the Bogoliubov
transformation and by diagonalization of the quantized cano\-ni\-cal
Hamilton equations in the Fock space we carry evolution from
operators onto the Bogoliubov coef\/f\/icients, \item We f\/ind
the f\/ield operator $\Psi$ of the Universe and conjugate momentum
operator $\Pi_\Psi$.
\end{enumerate}

\subsection{Quantum mechanics of the Universe}

In agreement with P.A.M. Dirac \cite{dir} we apply the f\/irst
quantization of the classical constraint equation. Recall that for
the Friedmann--Lema\^{i}tre Universe the Hamiltonian constraint
equation has a form
\begin{gather}\label{4.1}
p_{a}^{2}-{\omega}_{a}^{2}=0,
\end{gather}
with ${\omega}_{a}$ given by (\ref{3.10}). Classical solution of
this constraint equation is given by the Hubble law (\ref{3.14})
with the redshift (\ref{a1}).

The f\/irst quantization of the Hamiltonian constraint equation is
given by canonical commutation relation in the standard form
\[
i\left[\hat{\mathrm{p}}_{a},a\right]=1,
\]
where $\hat{\mathrm{p}}_{a}=-i\dfrac{\partial}{\partial a}$ is the
momentum operator corresponding to canonical conjugate
momentum~${p}_{a}$. We assume that the wave function $\Psi(a)$ for
quantum theory exists. The f\/inal result of this step is the
quantum evolution equation
\begin{gather}\label{4.3}
\left(\partial_{a}\partial_{a}+{\omega}_{a}^{2}\right)\Psi(a)=0,
\end{gather}
which is known as the Wheeler--DeWitt equation \cite{WDW,witt}.
This equation def\/ines quantum mechanics description of the
spacetime. Classical solutions of (\ref{4.3}) fulf\/ill all
conditions for wave function.

\subsection[Classical field theory of the Universe]{Classical f\/ield theory of the Universe}

The equation (\ref{4.3}) looks like the Klein--Gordon equation
\cite{pesk} for the boson with mass ${\omega}_{a}$. Let us
consider the Wheeler--DeWitt equation as an equation of motion for
the classical f\/ield $\Psi$ and describe the Hamiltonian
classical f\/ield theory of $\Psi$. For this purpose we should
construct the classical action which produces the equation of
motion (\ref{4.3}) from the Hamilton action principle. The correct
form of the classical action can be obtained by heuristic analogy
with the Klein--Gordon case
\begin{gather}\label{4.4}
\mathcal{S}[\Psi]=\dfrac{1}{2}\int_{a(\eta_0)}^{a(\eta)}
da\left\{\left(\partial_{a}\Psi\right)^2-{\omega}_{a}^{2}\Psi^{2}\right\}.
\end{gather}
Let us check that this action produces the Wheeler--DeWitt
equation. The Hamilton action principle gives
\begin{gather*}
\delta\mathcal{S}[\Psi]\equiv0 =
\dfrac{\delta\mathcal{S}[\Psi]}{\delta\Psi}\delta\Psi+\dfrac{\delta\mathcal{S}[\Psi]}
{\delta\partial_a\Psi}\delta\partial_a\Psi=\dfrac{\delta\mathcal{S}[\Psi]}{\delta\Psi}\delta\Psi
+\dfrac{\delta\mathcal{S}[\Psi]}{\delta\partial_a\Psi}\partial_a\delta\Psi\nonumber\\
\phantom{\delta\mathcal{S}[\Psi]\equiv0}{}=\left\{\dfrac{\delta\mathcal{S}[\Psi]}{\delta\Psi}-\partial_a\dfrac{\delta\mathcal{S}[\Psi]}{\delta\partial_a\Psi}\right\}\delta\Psi+\partial_a\left\{\dfrac{\delta\mathcal{S}[\Psi]}{\delta\partial_a\Psi}\delta\Psi\right\}.\nonumber
\end{gather*}
The second term vanishes on boundaries and thus  we obtain
\[
\dfrac{\delta\mathcal{S}[\Psi]}{\delta\Psi}-\partial_a\dfrac{\delta\mathcal{S}[\Psi]}{\delta\partial_a\Psi}=0,
\]
or after using (\ref{4.4})
\[
\int da
\left\{\omega_a^2\Psi+\partial_a\partial_a\Psi\right\}=0\Rightarrow\left(\partial_a\partial_a+\omega_a^2\right)\Psi=0,
\]
what is exactly the Wheeler--DeWitt equation (\ref{4.3}) for the
f\/ield $\Psi$. By this the Wheeler--DeWitt equation is an
equation of motion for the classical f\/ield $\Psi$ and the
heuristic action (\ref{4.4}) is correct.

Let us calculate the canonical conjugate momentum f\/ield
corresponding to the action (\ref{4.4})
\begin{gather}\label{4.5}
\Pi_{\Psi}=\dfrac{\delta\mathcal{S}[\Psi]}{\delta\left(\partial_{a}\Psi\right)}=\partial_{a}\Psi,
\end{gather}
With this momentum the action (\ref{4.4}) reduces into the form
\begin{gather}\label{4.6}
\mathcal{S}[\Psi]=\int{da}\left\{\Pi_{\Psi}\partial_{a}\Psi-\mathrm{H}(\Pi_{\Psi},\Psi)\right\},
\end{gather}
where
\begin{gather}\label{4.6a}
\mathrm{H}(\Pi_{\Psi},\Psi)=\dfrac{1}{2}\left(\Pi^{2}_{\Psi}+{\omega}_{a}^{2}\Psi^{2}\right),
\end{gather}
is the Hamiltonian that describes evolution of classical f\/ield
$\Psi$. The canonical Hamilton equations of motion for the
classical f\/ield theory described by the Hamiltonian (\ref{4.6a})
are as follows
\begin{gather}
\dfrac{\partial\mathrm{H}(\Pi_{\Psi},\Psi)}{\partial\Pi_{\Psi}}=\partial_{a}\Psi,\label{h1}\\
-\dfrac{\partial\mathrm{H}(\Pi_{\Psi},\Psi)}{\partial\Psi}=\partial_{a}\Pi_{\Psi}\label{h2},
\end{gather}
after calculations we can rewrite these equations in the form
\begin{gather}\label{h12}
\partial_{a}\left[\begin{array}{c}\Psi\\
\Pi_\Psi\end{array}\right]=\left[\begin{array}{cc}
0&1\\
-\omega_a^2&
0\end{array}\right]\left[\begin{array}{c}\Psi\\
\Pi_\Psi\end{array}\right].
\end{gather}
The equation (\ref{h1}) leads to the relation (\ref{4.5}), and the
equation (\ref{h2}) is equivalent to the Wheeler--DeWitt equation
(\ref{4.3}) after using the equation (\ref{h1}). Hitherto existing
approaches to quantization problem of the Friedmann--Lema\^{i}tre
spacetime, and generally to quantization problem for gravity, was
based on testing of solution or the second quantization of the
Wheeler--DeWitt equation for the theory. As opposed to these
approaches, we will base the quantum theory of gravity on the
canonical Hamilton equations of motion.

\subsection{Quantization of the Hamilton equations of motion}
The analogy method presented in previous subsection produces
conclusion that given system is a boson and by this the
corresponding quantum f\/ield theory description should be build
in boson type Fock space language. The boson type Fock space of
creation $\mathcal{G}^{\dagger}$ and annihilation $\mathcal{G}$
operators are standard, constructed by the canonical commutation
relations \cite{pesk,bog2,birula}
\begin{gather*}
\big[\mathcal{G}(a(\eta)),\mathcal{G}^{\dagger}(a(\eta'))\big]=\delta(a(\eta)-a(\eta')),
\\
\big[\mathcal{G}(a(\eta)),\mathcal{G}(a(\eta'))\big]=0.
\end{gather*}
By way of analogy with the Klein--Gordon theory we propose the
following second quantization  by the non-fockian type
distributions in the Fock space ($a\equiv a(\eta)$)
\begin{gather}
\Psi(a)=\dfrac{1}{\sqrt{2{\omega}_{a}}}\big(\mathcal{G}(a)+\mathcal{G}^{\dagger}(a)\big),\label{4.8}\\
\Pi_{\Psi}(a)=-i\sqrt{\dfrac{{\omega}_{a}}{2}}\big(\mathcal{G}(a)-\mathcal{G}^{\dagger}(a)\big),\label{4.9}
\end{gather}
or in compact form
\begin{gather}\label{h12x}
\left[\begin{array}{c}\Psi\\\Pi_\Psi\end{array}\right]
=\left[\begin{array}{cc}\dfrac{1}{\sqrt{2\omega_a}}&\dfrac{1}{\sqrt{2\omega_a}}\vspace{1mm}\\
-i\sqrt{\dfrac{\omega_a}{2}}&i\sqrt{\dfrac{\omega_a}{2}}\end{array}\right]
\left[\begin{array}{c}\mathcal{G}\\
\mathcal{G}^{\dagger}\end{array}\right].
\end{gather}
The correct canonical commutation relation for the f\/ield
operators $\Psi$ and $\Pi_\Psi$
\[
\left[\Psi(a(\eta')),\Pi_{\Psi}(a(\eta))\right]=i\delta(a(\eta)-a(\eta')),
\]
is preserved automatically. The distributions (\ref{4.8}) and
(\ref{4.9}) contain the new element that will play a crucial role
-- the normalization coef\/f\/icients depend on $a$. It causes that~(\ref{4.8}) and~(\ref{4.9}) are nonfockian representations in the
Fock space of the system.

Using the non-fockian distributions (\ref{4.8}) and (\ref{4.9}) we
can translate the Wheeler--DeWitt action (\ref{4.6}) into the Fock
space language
\[
\mathcal{S}(\mathcal{G},\mathcal{G}^{\dagger})=\int
\mathcal{D}a\left\{i\dfrac{\mathcal{G}^{\dagger}\partial_{a}\mathcal{G}-\mathcal{G}\partial_{a}\mathcal{G}^{\dagger}}{2}-\mathcal{H}\right\},
\]
where we have used the Feynman-type measure, and the ef\/fective
Hamiltonian $\mathcal{H}$ is equal to
\begin{gather}\label{4.11}
\mathcal{H}=\left(\mathcal{G}^{\dagger}\mathcal{G}+\dfrac{1}{2}\right){\omega}_{a}+\dfrac{i}{2}\big(\mathcal{G}^{\dagger}\mathcal{G}^{\dagger}
-\mathcal{G}\mathcal{G}\big)\Delta,
\end{gather}
where $\Delta=\dfrac{\partial_{a}{\omega}_{a}}{2{\omega}_{a}}$ has
the meaning of coupling. The Hamiltonian (\ref{4.11}) is well
known from the many particle theories as the Hamiltonian
describing \emph{the boson superfluidity phenomenon}. By this way
the coupling $\Delta$ manifests collective phenomena. The
superf\/luidity in quantum cosmology was first discussed in paper
\cite{PZ}.

\subsection{Diagonalization of equations of motion}
By quantization of the canonical Hamilton equations of motion
(\ref{h12}) we obtain the equations of motion for the creation and
annihilation operators in the Fock space
\begin{gather}\label{4.11a}
i\partial_{a}\left[\begin{array}{c}\mathcal{G}\\
\mathcal{G}^{\dagger}\end{array}\right]=\left[\begin{array}{cc}
-{\omega}_{a} & \fbox{$2i\Delta$}\vspace{1mm}\\
\fbox{$2i\Delta$}&
{\omega}_{a}\end{array}\right]\left[\begin{array}{c}\mathcal{G}\\
\mathcal{G}^{\dagger}\end{array}\right].
\end{gather}
These equations are understood as the Heisenberg equations for
$\mathcal{G}$ and $\mathcal{G}^{\dagger}$ \cite{bog2} with
nonlinearity in form of nondiagonal elements in the evolution
matrix (\ref{4.11a}). We see that the quantum evolution
(\ref{4.11a}) is not diagonal. Now we must diagonalize this
evolution. Firstly we use

{\bf The boson Bogoliubov transformation.} We change the basis
$(\mathcal{G}^{\dagger},\mathcal{G})$ to another basis
$(\mathcal{W}^{\dagger},\mathcal{W})$ in the Fock space by the
general transformation ($a\equiv a(\eta)$)
\[
\left[\begin{array}{cc}\mathcal{W}(a)\\
\mathcal{W}^{\dagger}(a)\end{array}\right]=\left[\begin{array}{cc}u(a)&v(a)\\
v^{\ast}(a)&u^{\ast}(a)\end{array}\right]\left[\begin{array}{cc}\mathcal{G}(a)\\
\mathcal{G}^{\dagger}(a)\end{array}\right].
\]
If we want to preserve the canonical commutation relations in the
basis $(\mathcal{W}^{\dagger},\mathcal{W})$
\[
\left[\mathcal{W}(a(\eta)),\mathcal{W}^{\dagger}(a(\eta'))\right]=\delta(a(\eta)-a(\eta')),
\qquad \left[\mathcal{W}(a(\eta)),\mathcal{W}(a(\eta'))\right]=0,
\]
we obtain the rotation condition
\begin{gather}\label{hip}|u(a)|^{2}-|v(a)|^{2}=1.
\end{gather}
After this we apply

{\bf Diagonalization of quantum canonical Hamilton equations of
motion.} The $a$-evolution in the basis
$(\mathcal{G}^{\dagger},\mathcal{G})$ (\ref{4.11a}) is transformed
into the evolution in the basis
$(\mathcal{W}^{\dagger},\mathcal{W})$ in the form
\begin{gather}\label{4.15}
i\partial_{a}\left[\begin{array}{c}\mathcal{W}\\
\mathcal{W}^{\dagger}\end{array}\right]=\left[\begin{array}{cc}
\omega_1 & 0 \\ 0 &
\omega_2\end{array}\right]\left[\begin{array}{c}\mathcal{W}\\
\mathcal{W}^{\dagger}\end{array}\right],
\end{gather}
with some diagonalization energies $\omega_1$ and $\omega_2$. In
this way there is no coupling in the basis
$(\mathcal{W}^{\dagger},\mathcal{W})$.

This procedure produces equations for the Bogoliubov
coef\/f\/icients $u$ and $v$
\begin{gather}\label{4.25}
i\partial_{a}\left[\begin{array}{c}v\\ u\end{array}\right]=\left[\begin{array}{cc}-{\omega}_{a}&-2i\Delta\\
-2i\Delta&{\omega}_{a}\end{array}\right]\left[\begin{array}{c}v\\u\end{array}\right].
\end{gather}
and values of the diagonalization energies $\omega_1$ and
$\omega_2$ are
\[
\omega_1=\omega_2=0\label{o2}.
\]
By this we have solution of the equations (\ref{4.15})
\[
\mathcal{W}(a)=\mathcal{W}(a_0), \qquad
\mathcal{W}^{\dagger}(a)=\mathcal{W}^{\dagger}(a_0),
\]
and we can see that the operator
$\mathcal{N}_\mathcal{W}=\mathcal{W}^{\dagger}\mathcal{W}=\mathcal{W}^{\dagger}(a_0)\mathcal{W}(a_0)$
is an integral of motion
\[
\partial_{a}\mathcal{N}_{\mathcal{W}}=0.
\]
By this the stable Bogoliubov vacuum state $|0\rangle$ exists
\[
\mathcal{W}|0\rangle=0, \qquad \langle0|\mathcal{W}^{\dagger}=0.
\]

Since the hyperbolic identity (\ref{hip}), the Bogoliubov
coef\/f\/icients $u$ and $v$ can be parameterized as \cite{ropke}
\[
v(a)=e^{i\theta(a)}\sinh \phi(a), \qquad u(a)=e^{i\theta(a)}\cosh
\phi(a),
\]
and thus  the equations (\ref{4.25}) are equivalent to the
equations
\[
\partial_a\theta(a)=\pm{\omega}_{a}=p_a,
\qquad
\partial_a\phi(a)=-2\Delta=-\dfrac{\partial_a\omega}{\omega}=-\partial_a\ln\left|\omega_a\right|,
\]
with obvious solutions
\[
\theta(a)=\int_{a_0}^{a}p_ada, \qquad
\phi(a)=-\ln{\left|\dfrac{{\omega}_{a}(\eta)}{{\omega}_{a}(\eta_0)}\right|}.
\]
By this we have
\begin{gather*}
v(a)=\dfrac{1}{2}\exp\left\{i\int_{a_0}^{a}p_ada\right\}\left(\dfrac{{\omega}_{a}(\eta_0)}{{\omega}_{a}(\eta)}-\dfrac{{\omega}_{a}(\eta)}{{\omega}_{a}(\eta_0)}\right),
\\
u(a)=\dfrac{1}{2}\exp\left\{i\int_{a_0}^{a}p_ada\right\}\left(\dfrac{{\omega}_{a}(\eta_0)}{{\omega}_{a}(\eta)}+\dfrac{{\omega}_{a}(\eta)}{{\omega}_{a}(\eta_0)}\right).
\end{gather*}

In the Einstein--Hilbert theory, gravitation does not exist
without structure of spacetime~-- the spacetime creates
gravitation, and gravitation creates the spacetime. The formalism
presented here describes the spacetime, which in our problem is
the Friedmann--Lema\^{i}tre Universe, in the language of
collective phenomena. In this formulation, these collective
phenomena take place in gas, which is a nontrivial mixture of
quanta of gravity and the quintessence, that is a model
approximation of bosons and fermions f\/ields. Generally our
proposition is based on applying of the graviton-matter gas
approach to quantization of gravity with matter f\/ields presence.
In this language we will formulate physics of the Universe.

\subsection{Field operator of the Universe}

As we have seen, the second quantization of the canonical Hamilton
equations of motion (\ref{h12}) really represents classical
f\/ield theory phase space $\left[\Psi(a)~ \Pi_\Psi(a)\right]^{T}$
by the non-fockian representation (\ref{h12x}) in the Fock space
with using of the correct Bogoliubov transformed basis
\[
\left[\begin{array}{cc}\mathcal{G}(a)\\
\mathcal{G}^{\dagger}(a)\end{array}\right]=\left[\begin{array}{cc}u^{\ast}(a)&-v(a)\\
-v^{\ast}(a)&u(a)\end{array}\right]\left[\begin{array}{cc}\mathcal{W}(a_0)\\
\mathcal{W}^{\dagger}(a_0)\end{array}\right].
\]
This procedure produces quantum f\/ield theory phase space
described by general relation
\[
\left[\begin{array}{c}\Psi(a)\\\Pi_\Psi(a)\end{array}\right]
=\left[\begin{array}{cc}\dfrac{u^{\ast}(a)-v^{\ast}(a)}{\sqrt{2\omega_a}}&\dfrac{u(a)-v(a)}{\sqrt{2\omega_a}}\vspace{1mm}\\
-i\sqrt{\dfrac{\omega_a}{2}}\left(u^{\ast}(a)+v^{\ast}(a)\right)&i\sqrt{\dfrac{\omega_a}{2}}\left(u(a)+v(a)\right)\end{array}\right]
\left[\begin{array}{cc}\mathcal{W}(a_0)\\
\mathcal{W}^{\dagger}(a_0)\end{array}\right],
\]
or after calculations
\[
\left[\begin{array}{c}\Psi(a)\\\Pi_\Psi(a)\end{array}\right]
=\left[\begin{array}{cc}\dfrac{1}{\omega_a(\eta_0)}\sqrt{\dfrac{\left|\partial_a\theta\right|}
{2}}e^{-i\theta}&\dfrac{1}{\omega_a(\eta_0)}\sqrt{\dfrac{\left|\partial_a\theta\right|}{2}}e^{i\theta}\vspace{1mm}\\
-i\omega_a(\eta_0)\sqrt{\dfrac{1}{2\left|\partial_a\theta\right|}}e^{-i\theta}&i\omega_a(\eta_0)\sqrt{\dfrac{1}{2\left|\partial_a\theta\right|}}e^{i\theta}\end{array}\right]
\left[\begin{array}{cc}\mathcal{W}(a_0)\\
\mathcal{W}^{\dagger}(a_0)\end{array}\right],
\]
where
\[
\partial_a\theta(a)=\pm\omega_a(\eta_0)\left(\dfrac{a(\eta)}{a(\eta_0)}\right)^2,
\qquad \theta(a)=\int_{a_0}^{a}p_ada.
\]

By this we have the f\/ield operator of the Universe
\[
\Psi(a)=\dfrac{1}{\omega_a(\eta_0)}\sqrt{\dfrac{\left|\partial_a\theta\right|}{2}}
\big(e^{i\theta}\mathcal{W}^{\dagger}(a_0)+e^{-i\theta}\mathcal{W}(a_0)\big),
\]
and similarly we can read the momentum f\/ield
\[
\Pi_\Psi(a)=i\omega_a(\eta_0)\sqrt{\dfrac{1}{2\left|\partial_a\theta\right|}}
\big(e^{i\theta}\mathcal{W}^{\dagger}(a_0)-e^{-i\theta}\mathcal{W}(a_0)\big).
\]

\section{\label{sec:8x}Thermodynamics of the Universe}

\subsection{Density matrix and entropy}
Obviously, the graviton-matter gas is an open quantum system
\cite{open} and should be described by nonequilibrium quantum
statistical mechanics methods \cite{ripka}. In the standard
approach to nonequilibrium processes the one-particle density
operator is the particle number operator. In the case of the
graviton-matter gas a role of particles is played by elements of
this gas. By this, the density operator for the system is
\[
\varrho_{\mathcal{G}}=\mathcal{G}^{\dagger}\mathcal{G},
\]
and if we rewrite this operator in
$(\mathcal{W},\mathcal{W}^{\dagger})$ basis we have
\[
\varrho_{\mathcal{G}}=\mathsf{W}^{\dagger}\rho\mathsf{W},
\]
where
$\mathsf{W}=\left[\begin{array}{c}\mathcal{W}\\\mathcal{W}^{\dagger}\end{array}\right]$
and
\[
\rho=\left[\begin{array}{cc}|u|^2&-uv\\-u^{\ast}v^{\ast}&|v|^2\end{array}\right],
\]
is the density matrix for graviton-matter gas in thermodynamical
equilibrium.

Physical entropy of the system is def\/ined by the formula well
known from the quantum information theory \cite{qit}
\[
\mathrm{S}=-\dfrac{\mathrm{tr}(\rho\ln\rho)}{\mathrm{tr}(\rho)}\equiv\ln\Omega,
\]
where $\Omega$ is the partition function that for the
graviton-quintessence gas is equal to
\begin{gather}\label{part}
\Omega=\dfrac{1}{2|u|^2-1}.
\end{gather}

\subsection{Temperature}
In case of the graviton-matter gas we have thermodynamical
nonequilibrium, particles of the gas go out from the system. As a
result we have diagonalized equations of motion, and we have found
basis where particle number operator is an integral of motion and
thus in this basis we have thermodynamical equilibrium of the
graviton-matter gas. So  we can use equilibrium statistical
mechanics formulas for thermodynamical description of the system.

If we identify the partition function of the graviton-quintessence
gas (\ref{part}) with the Bose--Einstein type partition function
we obtain
\[
\Omega=\dfrac{1}{2|u|^2-1}\equiv\dfrac{1}{\exp\dfrac{\mathrm{E}}{\mathrm{T}}-1}~~\Longrightarrow~~\mathrm{T}=\dfrac{\mathrm{E}}{\ln2|u|^2},
\]
where we used the Gibbs state type. This type of identif\/ication
has a meaning if and only if we identify
\[
\mathrm{E}\equiv\mathrm{U}-\mu\mathrm{N},
\]
where $\mathrm{U}$ is internal energy, $\mu$ is chemical potential
and $\mathrm{N}$ is number of particles of the graviton-matter
gas, respectively.

As we have seen, the Hamiltonian describes considered system was
given by (\ref{4.11})
\[%\label{eff}
\mathcal{H}=\left(\mathcal{G}^{\dagger}\mathcal{G}+\dfrac{1}{2}\right){\omega}_{a}
+\dfrac{i}{2}\big(\mathcal{G}^{\dagger}\mathcal{G}^{\dagger}-\mathcal{G}\mathcal{G}\big)\Delta.
\]
In a diagonalized basis this ef\/fective Hamiltonian has a form
\[
\mathcal{H}=\mathsf{W}^{\dagger}\mathbf{H}\mathsf{W},
\]
where
\[
\mathbf{H}=\left[\begin{array}{cc}\dfrac{|u|^2+|v|^2}{2}{\omega}_{a}
+i\dfrac{u^{\ast}v-uv^{\ast}}{2}\Delta&-uv{\omega}_{a}+i\dfrac{uu-vv}{2}\Delta\vspace{1mm}\\
-u^{\ast}v^{\ast}{\omega}_{a}+i\dfrac{v^{\ast}v^{\ast}-u^{\ast}u^{\ast}}{2}\Delta&\dfrac{|u|^2
+|v|^2}{2}{\omega}_{a}+i\dfrac{u^{\ast}v-uv^{\ast}}{2}\Delta\end{array}\right],
\]
is the matrix of the ef\/fective Hamiltonian.

In the quantum statistical mechanics \cite{ripka} internal energy
$\mathrm{U}$ of thermodynamical system is def\/ined by quantum
mechanical average Hamiltonian of the thermodynamical system
\[
\mathrm{U}=\langle\mathbf{H}\rangle
=\dfrac{\mathrm{tr}\,(\rho\mathbf{H})}{\mathrm{tr}\,\rho}.
\]
After averaging we obtain
\[
\mathrm{U}=\left(\dfrac{1}{2}+2\mathrm{N}+\dfrac{\mathrm{N}}{2\mathrm{N}+1}\right)\big(\sqrt{\mathrm{N}+1}
-\sqrt{\mathrm{N}}\big){\omega}_{a}(\eta_0),
\]
where $\mathrm{N}$ is a number of particles of the gas
\[
\mathrm{N}=|v|^2.
\]
So the chemical potential for the gas is
\[
\mu=\left[2+\dfrac{1}{(2\mathrm{N}+1)^2}-\dfrac{\dfrac{1}{2}+2\mathrm{N}+\dfrac{\mathrm{N}}{2\mathrm{N}+1}}{2\sqrt{\mathrm{N}(\mathrm{N}+1)}}\right]\left(\sqrt{\mathrm{N}+1}-\sqrt{\mathrm{N}}\right){\omega}_{a}(\eta_0),
\]
and temperature $\mathrm{T}$ (see Fig.~1) is equal to
\[
\mathrm{T}=\dfrac{\sqrt{\mathrm{N}+1}-\sqrt{\mathrm{N}}}{\ln(2\mathrm{N}+2)}\left[\left(\dfrac{1}{2}+2\mathrm{N}+\dfrac{\mathrm{N}}{2\mathrm{N}+1}\right)\left(1+\dfrac{1}{2}\sqrt{\dfrac{\mathrm{N}}{\mathrm{N}+1}}\right)-2\mathrm{N}-\dfrac{\mathrm{N}}{(2\mathrm{N}+1)^2}\right]{\omega}_{a}(\eta_0).
\]

\begin{figure}[t]
\centering
\begin{minipage}[t]{70mm}
\includegraphics[width=70mm]{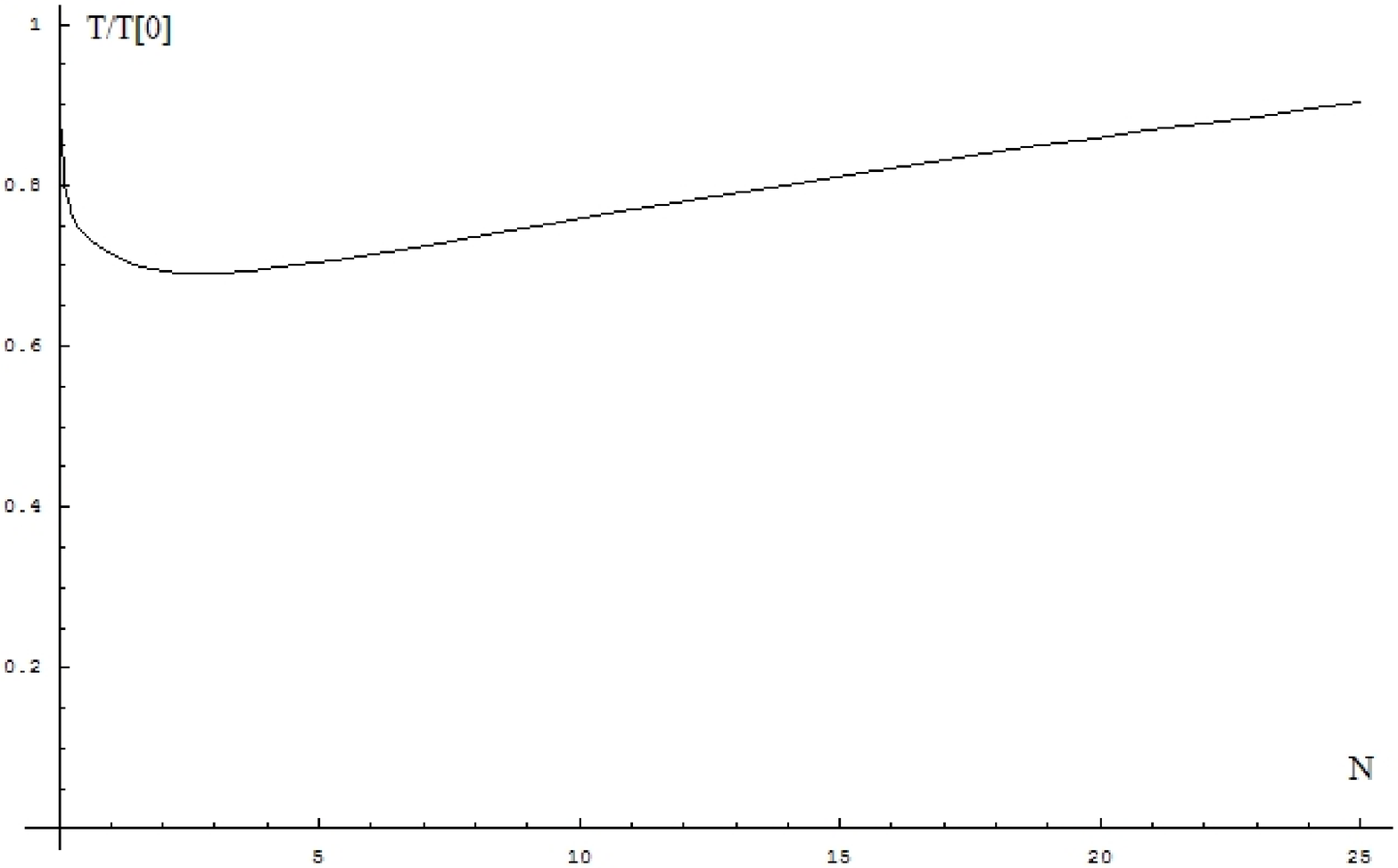}
\caption{Relation between temperature and number of particles for
the graviton-quintessence gas. Minimal value of temperature is
obtained for $N_0\approx2.73793853$ and is equal to
$\dfrac{T[N_0]}{T[0]}\approx0.69058084$.}
\end{minipage}\hfill
\begin{minipage}[t]{70mm}
\includegraphics[width=70mm]{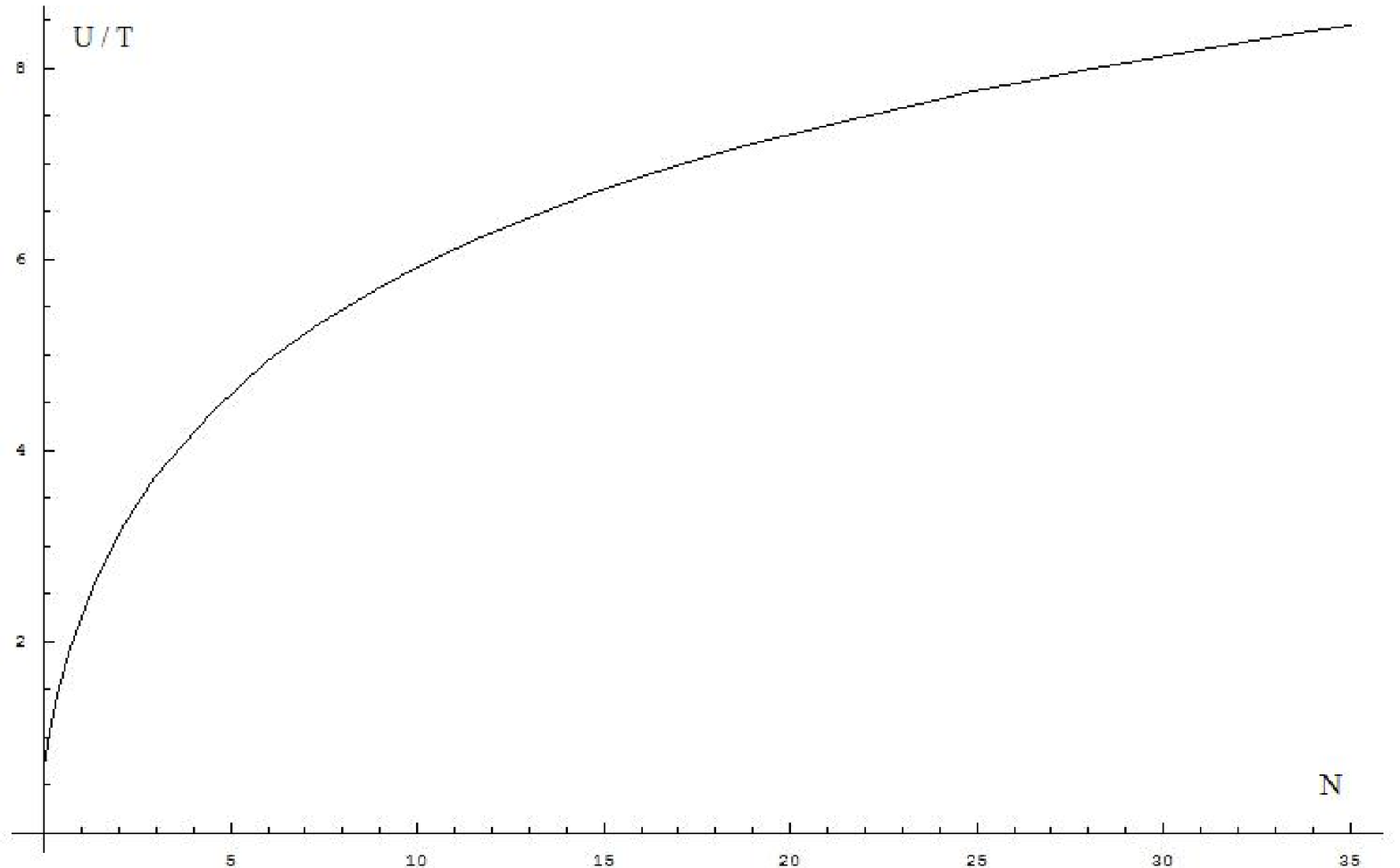}
\caption{The equation of state for the graviton-matter gas in case
of quintessence.}
\end{minipage}
\end{figure}

We see that now ($\mathrm{N}=0$) we have a f\/inite contribution
from the gas temperature
\[
\mathrm{T}[0]=\dfrac{{\omega}_{a}(\eta_0)}{\ln4}.
\]
Now we can conclude that the equation of state for the
graviton-matter gas in case of quintessence is equal to
\[
\dfrac{\mathrm{U}}{\mathrm{T}}=\dfrac{\ln(2\mathrm{N}+2)}{1+\dfrac{1}{2}\sqrt{\dfrac{\mathrm{N}}{\mathrm{N}+1}}-\dfrac{\mathrm{N}}{2\mathrm{N}+1}\dfrac{1+2(2\mathrm{N}+1)^2}{\mathrm{N}-1+3(2\mathrm{N}+1)^2}}.
\]
The diagram of this relation is presented on the Fig.~2.

\section{Graviton-matter gas as solution for quantum gravity}
In this paper we have considered the Friedmann--Lema\^{i}tre model
of the Universe with the quintessence. We have proposed the
quantization procedure for this classical cosmological model in
terms of the graviton-matter gas. As a result we have obtained
nontrivial formulation of cosmology in terms of collective
phenomena.

Physical meaning of the graviton-matter approach to the cosmic
microwave background radiation temperature anisotropies arises
from the following scenario. From the physical viewpoint we can
think about our Universe as a gas of gravitons, gauge bosons, and
material particles as electrons, quarks, Higgs particles etc. If
in our thinking huge volume of the Universe is taken into account,
the conclusion is that during our all observations and
measurements of the Universe physical properties, we are on the
position of an element of the gas~-- an observer in the Universe
is an element of the Universe. By this way observations of the
temperature anisotropies, understood as an ef\/fect of
condensation of all particles and f\/ields in the Universe, are
natural conceptual consequence of this approach. From the
graviton-matter gas viewpoint the quantum gravity has a meaning of
ef\/fective theory and collective phenomena language seems
adequate to description of the Universe physics. For this reason,
in our opinion, the graviton-matter gas approach is interesting
for further research in quantum cosmology.

\subsection*{Acknowledgements}
I am especially thankful to Victor N. Pervushin and Andrey B.
Arbuzov and for critical and develop remarks about my results. I
am grateful to Andrew Beckwith and Michel Vittot for their
interest in my solutions for quantum gravity.

\pdfbookmark[1]{References}{ref}
\LastPageEnding

\end{document}